\documentclass{mem}
\usepackage{natbib}\usepackage{txfonts}\usepackage{balance}
\usepackage{graphicx}
\usepackage[a4paper,breaklinks,dvipdfm]{hyperref}
\idline{75}{282}
\begin{document}
\def\teff{$T\rm_{eff }$}
\def\kms{$\mathrm {km s}^{-1}$}

\title{
Spectroscopic surveys of massive AGB and super-AGB stars
}

   \subtitle{}

\author{
D. A. Garc\'{\i}a-Hern\'andez\inst{1, 2}
          }

\institute{
Instituto de Astrof\'{\i}sica de Canarias, C/V\'{\i}a L\'actea s/n,
E$-$38205, Tenerife, Spain
\email{agarcia@iac.es}
\and
Departamento de Astrof\'{\i}sica, Universidad de La Laguna (ULL),
E-38206, Tenerife, Spain
}

\authorrunning{Garc\'{\i}a-Hern\'andez}

\titlerunning{Massive AGB and super-AGB surveys}

\abstract{
It is now about 30 years ago that photometric and spectroscopic surveys of
asymptotic giant branch (AGB) stars in the Magellanic Clouds (MCs) uncovered the
first examples of truly massive ($>$ 3$-$4 M$_\odot$) O-rich AGB stars
experiencing hot bottom burning (HBB). Massive (Li-rich) HBB AGB stars were
later identified in our own Galaxy and they pertain to the Galactic population
of obscured OH/IR stars. High-resolution optical spectroscopic surveys have
revealed the massive Galactic  AGB stars to be strongly enriched in Rb compared
to other nearby s-process elements like Zr, confirming that Ne$^{22}$ is the
dominant neutron source in these stars. Similar surveys of OH/IR stars in the
MCs disclosed their Rb-rich low-metallicity counterparts, showing that these
stars are usually brighter (because of HBB flux excess) than the standard
adopted luminosity limit for AGB stars (M$_{bol}$ $\sim$ $-$7.1) and that they
might have stellar masses of at least $\sim$6$-$7 M$_\odot$. The chemical
composition and photometric variability are efficient separating the massive AGB
stars from massive red supergiants (RSG) but the main difficulty is to
distinguish between massive AGB and super-AGB stars because the present
theoretical nucleosynthesis models predict both stars to be chemically
identical. Here I review the available multiwavelength (from the optical to the
far-IR) observations on massive AGB and super-AGB stars as well as the current
caveats and limitations in our undestanding of these stars. Finally, I underline
the expected observations on massive AGB and super-AGB stars from on-going
massive surveys like Gaia and SDSS-IV/APOGEE-2 and future facilities such as the
James Webb Space Telescope. 
}
\maketitle{}

\section{First identification of massive AGB stars and previous spectroscopic surveys}

The first identification of truly massive ($>$ 3$-$4 M$_\odot$) asymptotic giant
branch (AGB) stars dates back to about 30 years ago (Wood, Bessel \& Fox 1993).
Photometric surveys of AGB stars in the Magellanic Clouds (MCs) uncovered the
first examples of very luminous O-rich AGB stars, while the C-rich AGB stars are
generally found to be much fainter. These stars were found to be long-period
variables (of Mira type) with periods between $\sim$500 and 800 days and
enriched in heavy neutron-rich s-process elements. They were found to be
concentrated in a narrow luminosity range, with bolometric magnitudes
(M$_{bol}$) between $-$7 and $-$6, being consistent with relatively high-mass
progenitors (see Wood, Bessel \& Fox 1993, for more details). Subsequent
high-resolution optical spectroscopic surveys of visually bright AGB stars in
both MCs (LMC and SMC) discovered that these stars are rich in Li (see Figure
1), which confirmed the activation of the hot bottom burning (HBB) process in
these stars (see e.g., Smith \& Lambert 1989, 1990; Plez, Smith \& Lambert 1993;
Smith et al. 1995).

\begin{figure}[]
\resizebox{\hsize}{!}{\includegraphics[clip=true]{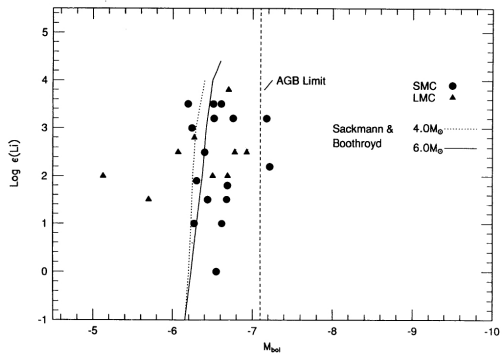}}
\caption{\footnotesize
Li abundances (derived by Smith et al. 1995) as a function of M$_{bol}$ in
O-rich HBB AGB stars in both Magellanic Clouds. Solar metallicitiy model
abundances for HBB from Sackmann \& Boothroyd (1992) are shown for two stellar
masses. Adapted from Smith et al. (1995).\label{eta}
}
\end{figure}

A more detailed chemical abundance analysis was carried out by Plez, Smith \&
Lambert in the early nineties. This study shows that the Li-rich HBB stars in
the SMC display very low C isotopic ratios (very near to the equilibrium values)
as expected from HBB models. However, these stars are not rich in Rb but rich in
other s-process elements like Zr and Nd (see Table 5 in Plez, Smith \& Lambert
1993). This suggests that these low-metallicity HBB stars produce s-process
elements via the $^{13}$C neutron source (see also Abia et al. 2001). More
recently, HBB stars have been identified in the very low-metallicity dwarf
galaxy IC 1613 (Menzies, Whitelock \& Feast 2015). One of these stars, G3011,
displays a strong Li line and it is very likely  Li-rich (see Fig. 2 in
Menzies, Whitelock \& Feast 2015). The average metallicity of IC 1613 is even
lower than the SMC, down to [Fe/H]$\sim$$-$1.6 dex, but the HBB stars are likely
younger and more metal-rich. 

In our own Galaxy, high-resolution optical spectroscopic surveys of very
luminous OH/IR stars were carried out about 10 years ago
(Garc\'{\i}a-Hern\'andez et al. 2006, 2007). Most of the stars with periods
longer than 400 days and OH expansion velocities (V$_{exp}$(OH)) higher than 6
kms$^{-1}$ were found to be Li-rich; with the Li abundances
(log$\varepsilon$(Li)) ranging from 1 to 3 dex, which confirm them as massive
HBB AGB stars. These massive Galactic AGB stars, however, are not rich in the
s-element Zr (Garc\'{\i}a-Hern\'andez et al. 2007). In strong contrast with the
SMC HBB stars, these Galactic stars displayed strong Rb overabundances; [Rb/Fe]
ranging from 0 to 2.6 dex (see Garc\'{\i}a-Hern\'andez et al. 2006). In short,
the more massive O-rich AGB stars of our Galaxy display strong Rb overabundances
with only mild Zr enhancements, as expected from the strong activation of the
Ne$^{22}$ neutron source (Garc\'{\i}a-Hern\'andez et al. 2006). 

Figure 2 displays the Rb abundances obtained by Garc\'{\i}a-Hern\'andez et al.
(2006) versus the V$_{exp}$(OH). The V$_{exp}$(OH) can be used as a
distance-independent mass indicator in OH/IR stars. The strong Rb enhancement
(up to $\sim$100 times solar) confirms the activation of the $^{22}$Ne neutron
source in massive AGBs. The AGB nucleosynthesis models can reproduce the
observed correlation between the Rb abundances and the stellar mass. However,
they cannot explain the extremely high Rb abundances seen in the more extreme
stars (Garc\'{\i}a-Hern\'andez et al. 2006).

\begin{figure}[]
\resizebox{\hsize}{!}{\includegraphics[clip=true,angle=-90]{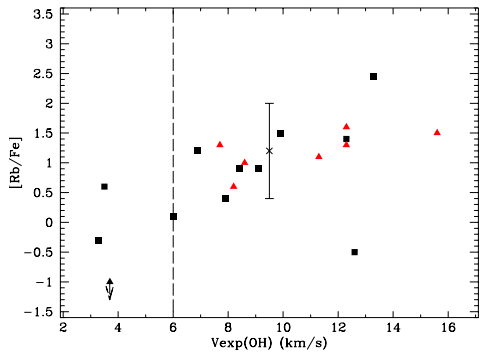}}
\caption{\footnotesize
Rb abundances in massive Galactic AGB stars, as obtained by
Garc\'{\i}a-Hern\'andez et al. (2006), versus OH expansion velocity
(V$_{exp}$(OH)). The  abundance estimates that correspond to the photospheric 
abundance needed to fit the stellar component are shown with red triangles. A
maximum error bar of $\pm$0.8 dex is also shown for comparison. The star with
high V$_{exp}$(OH) and  no  Rb, which  does  not  follow  the  correlation
observed is TZ Cyg, which possibly is a non-AGB star as it is not a long period,
Mira-like variable. Adapted from Garc\'{\i}a-Hern\'andez et al. (2006).
\label{eta2}}
\end{figure}

More recently, a few massive Galactic AGB stars at the beginning of the
thermally pulsing phase have been identified, permitting us to study the
nucleosynthesis at the early AGB stages (Garc\'{\i}a-Hern\'andez et al. 2013).
These stars are super Li-rich (log$\varepsilon$(Li) up to $\sim$4 dex) and the
strong Li is seen together with the complete lack of the s-process elements Rb,
Zr, and Tc, as predicted by the theoretical models. This confirms that HBB is
strongly activated at the early AGB stages and that the s-process is dominated
by the $^{22}$Ne neutron source.

\section{Metallicity effects}

The metallicity has a strong impact in the chemical evolution of massive AGB
stars. Table 1 summarizes the observational properties of massive AGB stars in
the Galaxy and the MCs. There are differences in the dust production, dredge-up
efficiency and AGB lifetime with metallicity. Also, the HBB is activated for
lower initial masses at lower metallicity. The Galactic stars are s-process poor
with very high Rb/Zr ratios typical of high neutron density and the $^{22}$Ne
neutron source. On the other hand, the MCs stars are rich in s-elements but
Rb-poor, with low Rb/Zr ratios characteristic of the lower neutron density of
the $^{13}$C neutron source.

\begin{table*}
\caption{Main observational properties of Galactic HBB AGB stars compared to
those of Magellanic Clouds (MCs) HBB AGB stars. Differences are attributed to
metallicity effects.}
\label{abun}
\begin{center}
\begin{tabular}{lccccccc}
\hline
\\
 & Dust        &           & AGB      & HBB        &        & Neutron  & Neutron \\ 
 & production  & Dredge-up & lifetime & activation & [s/Fe] & density  & source  \\
\hline
\\
Galaxy  & efficient     & non-efficient & small \# of TPs & for M$>$4M$_\odot$  & $<$0.5 dex & high & $^{22}$Ne \\
MCs     & non-efficient & efficient     & large \# of TPs & for M$>$3M$_\odot$  & $>$0.5 dex & low  & $^{13}$C 
\\
\hline
\end{tabular}
\end{center}
\end{table*}

But, where are the low-metallicity Rb-rich AGB counterparts? In 2009, we carried
out a high-resolution optical spectroscopic survey of luminous obscured O-rich
stars (including most of the known OH/IR stars) in the Magellanic Clouds and we
found the low-metallicity Rb-rich massive AGB counterparts
(Garc\'{\i}a-Hern\'andez et al. 2009). These stars display extremely high Rb
abundances (from $\sim$10$^{3}$ to 10$^{5}$ times solar). Interestingly, the Rb
abundance jumps at a M$_{bol}$ of $-$7 (see Fig. 3 in Garc\'{\i}a-Hern\'andez et
al. 2009); a result that it is very useful to distinguish such stars in other
Local Group galaxies. 

\section{Massive AGB or super-AGB stars?}

AGB evolutionary models for the LMC predict the very luminous massive AGB stars
mentioned above and the contribution of HBB to the luminosity explains their
luminosities in excess of the AGB theoretical limit. According to these models,
the Rb-rich AGBs in the LMC could have progenitor masses of at least 6$-$7
M$_\odot$ (see Fig. 7 in Ventura, D'Antona \& Mazzitelli 2000) so it is actually
not clear if they are massive AGB or super-AGB stars.

Indeed, both massive AGB and super-AGB stars can be easily separated from
massive red supergiants (RSGs) by using the photometric variability or the
chemical composition from high-resolution spectroscopy. However, the theoretical
nucleosynthesis models predict massive AGB and super-AGBs to be chemically
identical. In both types of stars, the HBB dominates and the s-process
nucleosynthesis is mainly driven by the $^{22}$Ne neutron source. Doherty et al.
(2017) present the s-process predictions for super-AGB stars at several
metallicities. It is clear that the s-process pattern of super-AGB stars is
identical to the one expected in massive AGB stars, with Rb being by far the
most abundant s-process element.

\section{The Rb problem}

The detection of Rb-rich AGB stars in the MCs (Garc\'{\i}a-Hern\'andez et al.
2009) uncovered a Rb problem that has two parts: i) the extremely high Rb
overabundances observed ([Rb/Fe]$\sim$3$-$5 dex); and ii) the large [Rb/Zr]
($\geq$3$-$4) ratios. The standard theoretical models (e.g., van Raai et al.
2012) are far from matching the observational results and within the framework
of the s-process it is not possible to overproduce Rb without co-producing Zr at
similar levels.

These standard theoretical models qualitatively describe the observations of
Rb-rich AGB stars in both the MCs and our Galaxy, in the sense that increasing
Rb abundances with increasing stellar mass and decreasing metallicity are
theoretically predicted (see Table 2 in Garc\'{\i}a-Hern\'andez et al. 2009).
However, the Rb abundance and the [Rb/Zr] ratio do not reach extreme values.

The extremely high Rb enhancements and  [Rb/Zr] ratios are likely artefacts of
the abundance analysis. So, the adopted hydrostatic model atmospheres likely
fail to represent the real stars. More realistic model atmospheres for extreme
AGB stars have been developed; i.e., taking into account the presence of a
circumstellar envelope with a radial wind (Zamora et al. 2014). The main result
of these new pseudo-dynamical (extended atmosphere) models is that the effect of
the circumstellar envelope is dramatic and the new Rb abundances are much lower
(sometimes by orders of magnitude) than those obtained with classical
hydrostatic models (e.g., Garc\'{\i}a-Hern\'andez et al. 2006, 2009). On the
other hand, Zr is practically non affected by the presence of the circumstellar
envelope and the Zr abundances with pesudo-dynamical models are nearly solar;
i.e., very similar to those from the hydrostatic ones. This is because the ZrO
bandhead used in the chemical analysis is formed deeper in the atmosphere and
much less affected than Rb (see Zamora et al. 2014). 

The Rb abundances in the full sample of massive AGB stars in our Galaxy have
been recently re-calculated with these extended atmosphere models by
P\'erez-Mesa et al. (2017). The results are shown in Figure 3, which compares
the abundances with hydrostatic models with the new ones with pseudo-dynamical
models. An important result is that the Rb abundances strongly depend on the
mass-loss rate, which is unknown for these stars. Thus, in order to break the
degeneracy in the model fits, accurate mass-loss rate estimates in these stars
(e.g., by observing the CO rotational lines in the radio domain) would be
needed.

\begin{figure*}[tbh]
\resizebox{0.90\hsize}{!}{\includegraphics[clip=true]{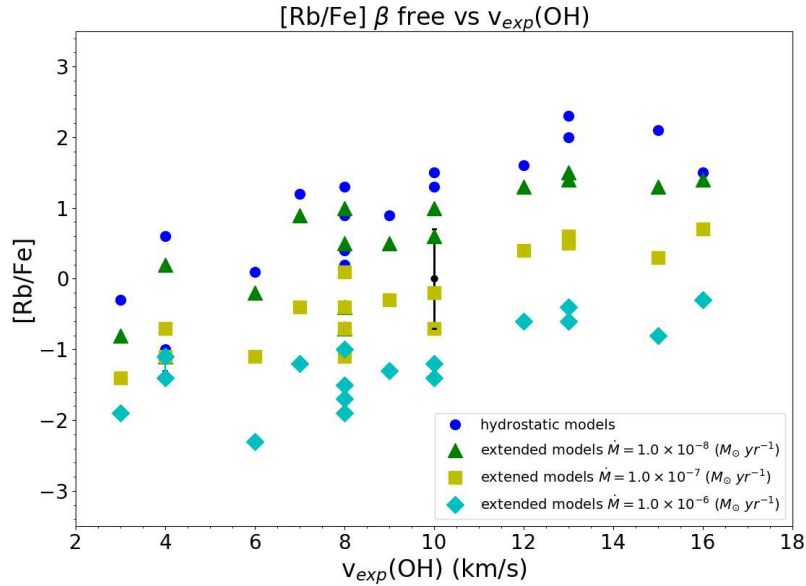}}
\caption{\footnotesize
Rb abundances vs. the expansion velocity (v$_{exp}$(OH)) for extended model
atmospheres with $\dot{M}$ = 10$^{-8}$, 10$^{-7}$ and 10$^{-6}$
M$_{\odot}yr^{-1}$ (green triangles, yellow squares and cyan diamonds,
respectively), as obtained by P\'erez-Mesa et al. (2017), in comparison with
those obtained by Garc\'{\i}a-Hern\'andez et al. (2006) from hydrostatic models
(blue dots). Adapted from P\'erez-Mesa et al. (2017).
\label{eta3}}
\end{figure*}

In short, the new Rb abundances and [Rb/Zr] ratios derived with more realistic
AGB model atmospheres significantly resolve the problem of the mismatch between
the observations of massive Rb-rich AGB stars and the theoretical predictions.
The new [Rb/Fe] abundances and [Rb/Zr] ratios range from 0.0 to 1.3 dex and from
$-$0.3 to $+$0.7, respectively (see Fig. 11 in P\'erez-Mesa et al. 2017). These
ranges are in good agreement with the synthetic results of the Monash models;
both the standard (van Raai et al. 2012; Karakas \& Lugaro 2016) and the delayed
superwind models (Karakas, Garc\'{\i}a-Hern\'andez \& Lugaro 2012). However, the
FRUITY\footnote{FUll-Network Repository of Updated Isotopic Tables and Yields:
http:// fruity.oa-teramo.inaf.it/.} AGB models predict too low Rb abundances,
being at odds with the observational evidence.  

Also, we very recently re-calculated the abundances of Li by using these
pseudo-dynamical models (P\'erez-Mesa et al.; these proceedings). The main
finding is that the Li abundances are only slightly affected by circumstellar
effects and the Li abundances are practically identical to those previously
obtained with hydrostatic models, which further confirm the HBB activation in
massive AGBs of our Galaxy. In this case, the much lower Li column density as
compared to Rb, likely explains the results with extended atmosphere models.

Finally, we note that the far-IR Herschel observations of extreme Galactic OH/IR
stars also confirm the HBB nature of these stars. The O and C isotopic ratios
obtained from the observed water and CO lines are consistent with HBB (see e.g.,
Justtanont et al.; these proceedings).

\section{On-going spectroscopic surveys and future facilities}

The SDSS-IV/APOGEE-2 is an on-going massive spectroscopic survey of Galactic
giant stars in the near-IR H-band (see Blanton et al. 2017 and references
therein). The resolution is about 20,000 and the H-band permits to obtain the
abundances of up to 15 elements by using molecular (e.g., OH, CO and CN) and
atomic lines. The abundances of some s-process elements like Nd and Ce as well
as the C isotopic ratios can be also obtained in an important fraction of the
stars observed. The APOGEE-2 survey is the continuation of APOGEE-1, which
already observed more than 150,000 stars. It is composed by two identical
instruments in both hemispheres. APOGEE-2 will observe more than 300,000 stars
(mainly RGB and AGB stars). Only a few detailed chemical analysis (from
high-resolution near-IR spectroscopic data) of massive AGB stars have been
carried out so far (e.g., McSaveney et al. 2007). Thus, APOGEE-2 will permit
better studies of the HBB, third dredge-up and the s-process in massive AGB
stars by observing a larger number of stars; especially in the inner Galaxy and
the MCs. New discoveries are also expected; e.g., the very recent identification
of very low-metallicity HBB-AGB stars towards the Galactic bulge (Zamora et al.,
in preparation). 

In addition, we are now in the Gaia era. Gaia is expected to provide the
distances (and so the luminosities) for all types of Galactic AGB stars, giving
strong constraints to the AGB theoretical models. For example, the luminosities
would help to better distinguish massive AGBs from the more luminous super-AGB
stars. However, we anticipate that this is going to be rather difficult because
these stars are very faint (and strongly variable) in the optical Gaia bands. 

Finally, the future James Webb Space Telescope (JWST) will be much more
sensitive than Spitzer, with access to the near-IR range. This mission will
permit near- and mid-IR spectroscopic studies of individual massive AGB and
super-AGB stars in Local Group galaxies. Such observations would permit to study
their dust chemical composition, mass-loss rates, dust production, as well as
their temperatures and possibly abundances (see Boyer; these proceedings).

\section{Summary}

In summary, multiwavelength (from the optical to the far-IR) spectroscopic
observations confirm the HBB and $^{22}$Ne activation in massive AGBs (and
super-AGB stars). Accurate mass-loss rates in these stars can break the
degeneracy of the more realistic extended model atmospheres developed for these
stars, which will permit more reliable Rb abundances. At present, we can
efficiently distinguish between massive AGBs and RSGs; by using the photometric
variability information and the chemical composition from high-resolution
spectroscopic observations. However, massive AGB stars are chemically identical
to super-AGBs and actual observational samples may contain super-AGB stars.
On-going massive spectroscopic (and photometric) surveys like SDSS-IV/APOGEE-2
and Gaia as well as future facilities like the JWST will contribute to a better
understanding of massive AGB stars and perhaps to the first unambiguous
detection of a super-AGB star.
 
\begin{acknowledgements}
DAGH acknowledges support provided by the Spanish Ministry of Economy and
Competitiveness (MINECO) under grant AYA$-$2014$-$58082$-$P. DAGH was also
funded by the Ram\'on y Cajal fellowship number RYC$-$2013$-$14182.
\end{acknowledgements}

\bibliographystyle{aa}

\end{document}